\documentclass{article}

 \usepackage[preprint]{neurips_2025}
\setcitestyle{numbers,square}

\usepackage[utf8]{inputenc} 
\usepackage[T1]{fontenc}    
\usepackage{hyperref}       
\usepackage{url}            
\usepackage{booktabs}       
\usepackage{amsfonts}       
\usepackage{nicefrac}       
\usepackage{microtype}      
\usepackage{xcolor}         

\usepackage[most]{tcolorbox}
\usepackage{amsmath}
\usepackage{geometry}
\usepackage{float}

\title{ClimateAgents: A Multi-Agent Research Assistant for Social-Climate Dynamics Analysis}

\author{
	Shan Shan \\
	Department of Mathematics\\
	 International Center for Interdisciplinary Statistics\\
	Harbin Institute of Technology\\
 \\
	\texttt{shans@hit.edu.cn} \\
}

\begin{document}

\maketitle

\begin{abstract}

The complex interaction between social behaviors and climate change requires more than traditional data-driven prediction; it demands interpretable and adaptive analytical frameworks capable of integrating heterogeneous sources of knowledge. This study introduces \textit{ClimateAgents}, a multi-agent research assistant designed to support social-climate analysis through coordinated AI agents. Rather than focusing solely on predictive modeling, the framework assists researchers in exploring socio-environmental dynamics by integrating multimodal data retrieval, statistical modeling, textual analysis, and automated reasoning. Traditional approaches to climate analysis often address narrowly defined indicators and lack the flexibility to incorporate cross-domain socio-economic knowledge or adapt to evolving research questions. To address these limitations, \textit{ClimateAgents} employs a set of collaborative, domain-specialized agents that collectively perform key stages of the research workflow, including hypothesis generation, data analysis, evidence retrieval, and structured reporting. The framework supports exploratory analysis and scenario investigation using datasets from sources such as the United Nations and the World Bank. By combining agent-based reasoning with quantitative analysis of socio-economic behavioral dynamics, \textit{ClimateAgents} enables adaptive and interpretable exploration of relationships between climate indicators, social variables, and environmental outcomes. The results illustrate how multi-agent AI systems can augment analytical reasoning and facilitate interdisciplinary, data-driven investigation of complex socio-environmental systems.

\end{abstract}

\section{Introduction}

When dealing with complex problems, people often break them into smaller parts and coordinate these parts to reach a solution. Advances in automation have made this process smoother and more efficient by allowing different tasks to be organized and carried out within one system. Recent developments in AI agents extend this idea. In agent-based systems, tools, models, and data access can be wrapped into specialized agents that work together to solve complex problems. In this sense, \textbf{an effective research assistant performs a similar role: helping gather information, analyze data, explore ideas, and organize results while supporting the researcher’s decision-making process.}

Social behavior and climate change are examples of such complex domains. They involve interactions across social, economic, and environmental factors. Studying these problems requires frameworks that can combine different sources of knowledge and support multiple types of analysis. Using the ideas of automation together with recent advances in AI agents provides a practical way to study these social-climate dynamic systems in a more scalable and flexible way. 

This raises an important question: \textbf{how can we build a framework that truly functions as a research assistant for complex socio-environmental analysis? }Motivated by practical research experience, this paper proposes \textit{ClimateAgents}.

Specifically, social behavior is a key driver of climate change, yet current modeling approaches often struggle to capture its complexity across spatial, temporal, and disciplinary scales. Particularly, data-driven models tend to prioritize narrowly defined climate objectives, offering limited flexibility to incorporate out-of-domain socio-economic knowledge and constrained adaptability to emerging societal challenges. In response to these limitations, this study introduces \textit{ClimateAgents}, a generative AI framework that embeds large language model (LLM)-powered AI Agents' reasoning within a system of domain-specialized agents. The framework supports multimodal data retrieval, statistical simulation, and dynamic interpretation across social and economic dimensions, thereby enhancing the scope and depth of climate policy analysis. Specifically, ClimateAgents enables exploratory hypothesis generation, scalable textual analysis, and scenario construction using data from sources such as the United Nations and the World Bank.

Technically, the architecture draws inspiration from Marvin Minsky’s theory of the mind, as articulated in \textit{The Society of Mind} \cite{minsky1988society}, where intelligence emerges from the interaction of numerous simple agents. According to this view, each agent represents a functional unit with limited capability, but through their organized interplay, complex cognitive behaviors arise. Intelligence, therefore, is not a property of any single part, but rather an emergent phenomenon of the whole.

Recent advances in AI agents have expanded the capabilities of artificial intelligence systems across scientific and engineering domains. Rather than functioning as standalone predictive models, modern LLM-based agents integrate reasoning, tool use, memory, and external knowledge sources to perform complex tasks through iterative planning and interaction with computational environments. This shift toward agentic AI has enabled systems capable of autonomous task decomposition, hypothesis generation, and long-horizon decision-making workflows. Examples include software engineering agents such as SWE-Agent and execution-environment systems like OpenHands that allow language models to interact directly with development environments and external tools \cite{yang2024sweagent,openhands2024}. These developments demonstrate how AI agents can operate beyond static text generation and instead perform actions within real computational systems.

More recently, a new generation of open-source autonomous agent frameworks has emerged, further expanding the practical capabilities of agentic systems. Platforms such as OpenClaw provide self-hosted AI agents capable of executing multi-step workflows, interacting with applications, and controlling local computing environments through messaging interfaces and programmable tool ecosystems. These agents can automate tasks such as email management, code execution, and web interaction while maintaining persistent memory and external tool access, enabling continuous operation beyond single-turn conversations \cite{chen2026openclaw, wang2026openclawrl}. Empirical studies further suggest that large-scale communities of autonomous agents can emerge on dedicated platforms where agents interact and learn within agent-native social environments \cite{chen2026openclaw}. Such systems illustrate the growing transition from single-model AI assistants toward distributed agent ecosystems capable of coordinated activity and persistent interaction.

Despite these advances, significant challenges remain when applying LLM-based agents to the social sciences. Social systems are inherently dynamic and involve multiple actors whose behaviors evolve through interaction, feedback, and shifting contextual conditions. Current agentic systems still struggle to faithfully represent these multi-actor dynamics, maintain accurate situational awareness in rapidly evolving environments, and integrate continuously updated information beyond their static training data. These limitations constrain the ability of existing AI agents to model complex socio-technical systems characterized by emergent behavior and interdependent decision processes.

Recent AI agent frameworks have  advanced the development of multi-agent systems built on large language models. Platforms such as AutoGen \cite{wu2024autogen}, CrewAI \cite{moura2024crewai}, LangGraph \cite{langgraph2024}, and BeeAI \cite{beeai2024} provide pipeline-based architectures for coordinating interactions among agents and external tools. These systems are effective for workflow automation and collaborative reasoning tasks. Other frameworks, including CAMEL \cite{li2023camel}, AgentScope \cite{gao2024agentscope}, AgentVerse \cite{chen2024agentverse}, and AgentSociety \cite{li2025agentsociety}, explore large-scale societies of language agents and emergent multi-agent behaviors. While these platforms support experimentation with agent communication and coordination, they primarily focus on general-purpose agent interactions rather than domain-specific simulation environments.

To address these challenges and extend the applicability of AI agents to complex social systems, this work proposes a multi-agent architecture inspired by Minsky's \emph{Society of Mind} framework \cite{minsky1988society}. Rather than treating the AI agent as a monolithic intelligence, the proposed framework distributes reasoning and task execution across a set of interacting agents, each responsible for specialized functions. The architecture consists of three core layers: (i) the \textbf{Perception Layer}, which processes multimodal inputs—including text, tables, and images—into structured representations; (ii) the \textbf{Reasoning Layer}, powered by an LLM-based agent responsible for planning, inference, and coordination; and (iii) the \textbf{Operation Layer}, which executes reasoning-driven actions such as data retrieval, statistical analysis, and visualization through external tools and computational modules. This layered, agent-oriented design supports iterative feedback, coordination, and adaptive problem solving in complex environments.

The structure of this paper is as follows. Section1 details the design of the multi-agent system and its component layers. This is followed by a series of computational experiments that demonstrate the effectiveness of multi-agent collaboration in complex tasks. Section 2 concludes with a discussion of limitations, open challenges, and future research directions.

Informed by Minsky’s philosophy, the study focuses on the socio-behavioral dimensions of climate action, with particular emphasis on United Nations Sustainable Development Goal 13. The contributions of this work are threefold: (i) an interpretable and adaptive alternative to predictive-only models; (ii) an integrated methodology for aligning AI-generated insights with human behavioral contexts; and (iii) a tool for interdisciplinary climate research that bridges empirical analysis and policy relevance. The results illustrate new opportunities for designing climate policy interventions grounded in social insight and computational reasoning.

\begin{figure}[ht]
	\centering
	\begin{minipage}[b]{0.49\linewidth}
		\includegraphics[width=\linewidth]{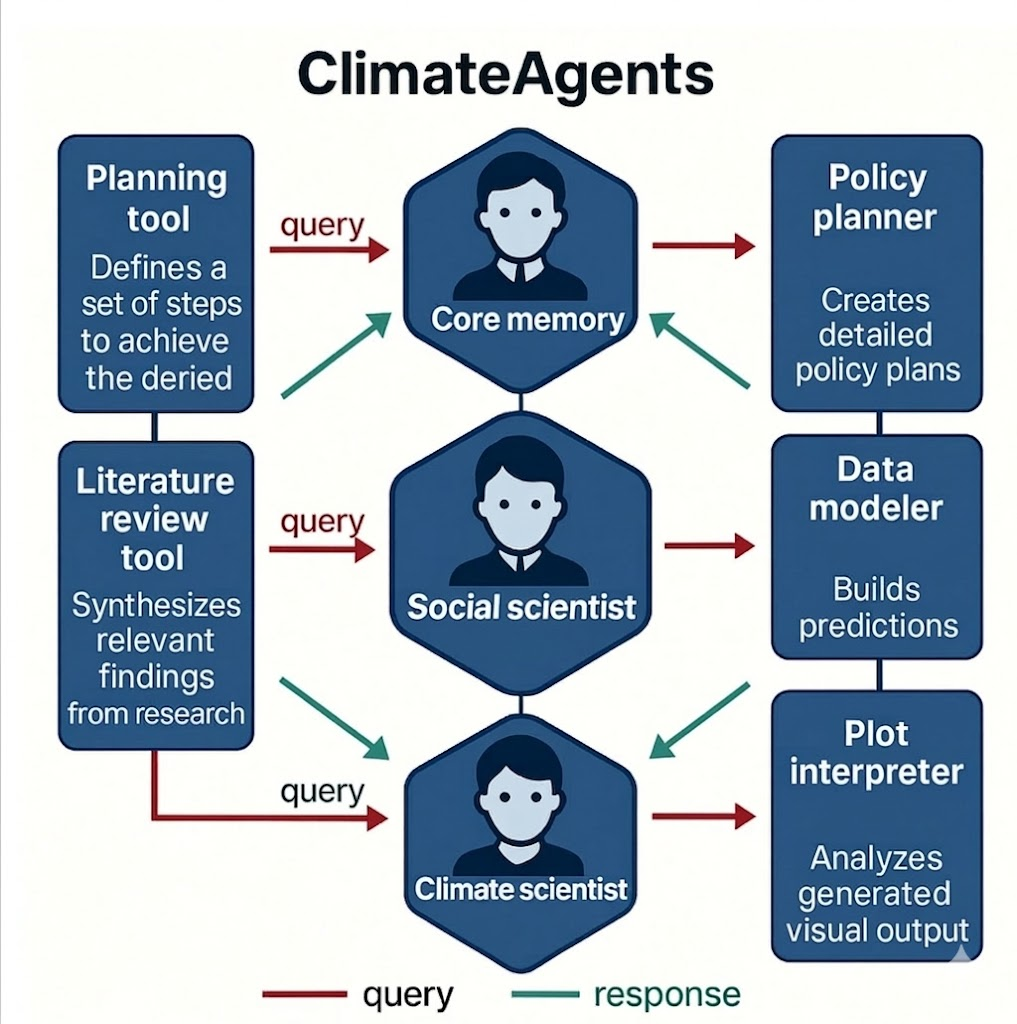}
		\caption{The ClimateAgents LLM generating a sequence of "mixed" and "why" questions to probe the causal and correlative relationships among social indicators and carbon emissions.}
		\label{fig:left}
	\end{minipage}
	\hfill
	\begin{minipage}[b]{0.49\linewidth}
		\includegraphics[width=\linewidth]{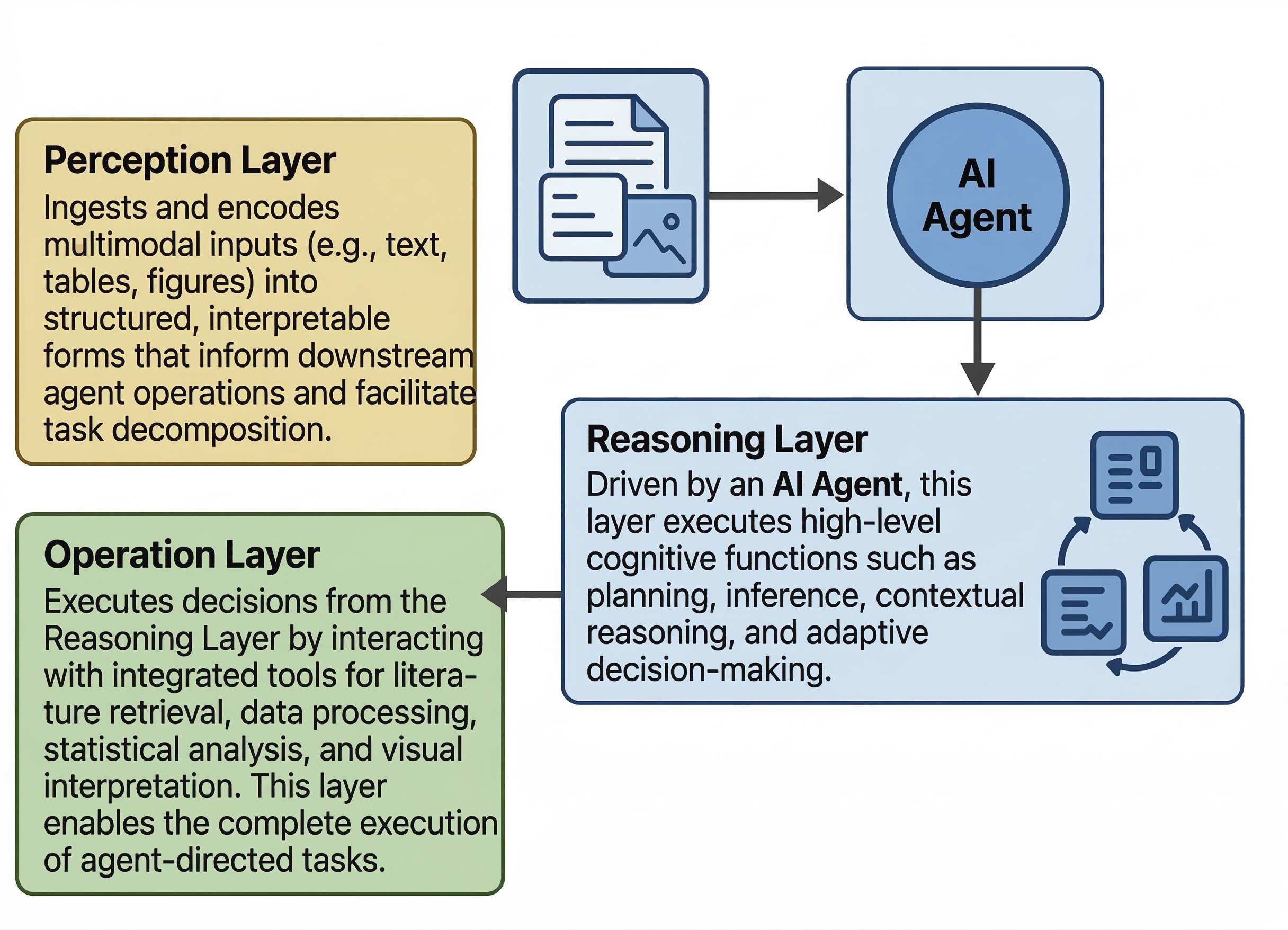}
		\caption{A comparative analysis between user-supplied and LLM-generated responses to climate-related prompts, highlighting differences in interpretability and reasoning.}
		\label{fig:right}
	\end{minipage}
\end{figure}

\section{Related works}

Advancements in prompt engineering and multimodal generative AI have jointly expanded the capabilities of language and vision models, especially in domains like healthcare, design, and scientific inquiry. Engineering effective prompts is central to harnessing these systems' potential, particularly in zero-shot and few-shot contexts. Strategies such as the Zero-shot Evaluation Instruction and Zero-shot-CoT Instruction enable models to generalize without prior examples, the latter invoking intermediate reasoning steps-known as chain-of-thought prompting-to improve task performance \citep{Brown2020, zhou2022large, Srivastava2022}. In contrast, Few-shot Evaluation and Resample Instructions leverage minimal examples or iterative feedback to refine model outputs, while Forward Generation predicts subsequent content based on context, supporting more dynamic natural language generation.
\footnote{\url{https://platform.openai.com/docs/guides/prompt-engineering} \ \url{https://huggingface.co/spaces/Gustavosta/MagicPrompt-Stable-Diffusion} \ \url{https://promptomania.com/stable-diffusion-prompt-builder/}}

These prompt strategies are increasingly relevant as multimodal generative AI systems-those integrating visual, textual, and sometimes auditory modalities-become more sophisticated. For instance, \citet{lu2024multimodal} introduced a vision-language assistant for pathology diagnostics that leverages GPT-4V to match expert clinical performance. Broadening this, \citet{rao2025multimodal} surveyed the landscape of multimodal AI in medical image interpretation, highlighting how cross-modal data integration enhances clinical decision-making. 

From a technical perspective, \citet{zhan2023multimodal} reviewed the core challenges in multimodal image synthesis and editing, which underlie the alignment mechanisms used in prompt-guided generation. Likewise, \citet{poon2023multimodal} emphasized the utility of multimodal AI in precision health, particularly through the integration of imaging and multi-omics data. Structural underpinnings of these models-often based on latent variable architectures-were examined in \citet{suzuki2022survey}, while \citet{rasiklal2024review} offered a comprehensive overview of multimodal integration spanning text, image, and audio. In design contexts, \citet{peng2024designprompt} demonstrated how multimodal prompt input can stimulate creative exploration, showcasing prompt engineering not just as a technical method but also as a cognitive tool. Beyond practical applications, multimodal generative AI also raises epistemological considerations. \citet{cope2024grammar} analyzed how such technologies reshape communication and meaning-making, while \citet{alasadi2024chemistry} illustrated their utility in scientific domains such as visual problem-solving in chemistry.

Despite these advances, there remains a notable gap in datasets tailored for causal reasoning. While general-purpose datasets from sources like Google \citep{kwiatkowski2019natural}, Bing \citep{nguyen2016ms}, and LLM-user interactions (e.g., ShareGPT and WildChat) \citep{zhao2024wildchat} exist, few directly address causal inquiry, especially in prompts designed for LLMs \citep{ouyang2022training, jin2024largelang}. There is also no existing dataset focusing on climate-related causal questions, underscoring a crucial area for future dataset development \citep{ceraolo2024}.

However, all of the aforementioned components-while impressive in isolation-still fall short of embodying what thinkers like Minsky \cite{minsky1986society,minsky1988society,minsky2007emotion} and Alan Turing \cite{turing1950mind} envisioned as true intelligence. These advances represent fragmented capabilities, each addressing a specific cognitive function. But what if we move beyond assembling isolated parts? What if we aim to integrate them holistically within a social context, where intelligence is not just individual but interactive, adaptive, and grounded in real-world human dynamics? 

As illustrated at the beginning, social domains are inherently complex and diverse. When it comes to a global challenge like climate change, the question becomes: how can we leverage the strengths of existing AI components-each excelling in specific tasks-and pipeline them into a unified, intelligent, and holistic agent? Such an agent would not only process multimodal information but also reason causally, interact socially, and adapt dynamically to the nuances of real-world decision-making within the climate domain.

\section{Methods}

To investigate the capabilities of AI agents in climate and scientific reasoning tasks, this work introduces a multi-agent framework- \textit{ClimateAgents}- inspired by Marvin Minsky’s theory of the mind, wherein intelligent behavior emerges from the coordinated activity of numerous simple agents. In alignment with this perspective, complex tasks are decomposed into modular cognitive processes distributed across specialized agents.

As illustrated in Figure~\ref{fig:left}, the ClimateAgents LLM generates structured sequences of ``mixed'' and ``why'' questions designed to uncover causal and correlative relationships between social indicators and carbon emissions. These questions support exploratory reasoning and hypothesis development. Figure~\ref{fig:right} presents a comparison between user-generated and model-generated responses to climate-related prompts, highlighting distinctions in interpretability and reasoning depth.

To overcome the limitations of static, single-turn LLM interactions, a \textbf{multi-agent system architecture} is proposed to extend the functional range of LLMs through a layered design. This architecture comprises three primary components: (i) \textbf{Perception Layer}, responsible for processing multimodal inputs-including text, structured tables, and visual data-into interpretable representations for downstream agents; (ii) \textbf{Reasoning Layer}, powered by a frontier LLM, which performs high-level cognitive operations such as inference, planning, and decision-making; and (iii) \textbf{Operation Layer}, which executes actions informed by the reasoning process, such as information retrieval, statistical analysis, and data visualization. This structure enables continuous feedback and iterative refinement through external tool integration.

The proposed framework facilitates dynamic integration of heterogeneous data sources, scalable task decomposition, and coordinated agent collaboration. While demonstrated within the context of climate indicator analysis, the architecture is intended to generalize across other scientific domains-such as materials design-where conventional human-centric methods often encounter challenges related to complexity and data heterogeneity.

AI agents are designed using state-of-the-art, general-purpose large language models from the GPT-4 family. Dynamic multi-agent collaboration is implemented within the AutoGen framework, an open-source ecosystem for agent-based AI modeling. In the ClimateAgents multi-agent system, the \textit{user}, \textit{admin}, and \textit{executor} agents are instantiated using the \texttt{UserProxyAgent} class from AutoGen. Agents supporting the “Knowledge Retrieval” tool-namely, the \textit{assistant} and \textit{reviewer}—are created using the \texttt{RetrieveAssistantAgent} class. The “Plot Analyzer” agent, responsible for interpreting and critiquing data visualizations, is implemented via the \texttt{MultimodalConversableAgent} class. Remaining agents are instantiated through the \texttt{AssistantAgent} class, while coordination and task delegation across agents are managed using the \texttt{GroupChatManager} class. Each agent is assigned a well-defined role, expressed through a profile description embedded in its system message at creation. This role-based architecture enables modular, context-sensitive task execution across diverse subtasks. Further implementation details are available in the source code repository hosted on GitHub.

\textbf{ClimateAgents} is designed as a generative, multi-agent model tailored for addressing complex tasks in social climate research. The system architecture consists of a core team of collaborating agents equipped to analyze multidimensional climate indicators-such as emissions, policy interventions, social equity, and economic factors-using a suite of integrated tools. These tools include modules for natural language retrieval, computational simulation, policy document parsing, and visual data interpretation. Each tool is composed of multiple AI agents with specific competencies, working collaboratively to respond to user queries related to climate action and societal behavior. Agent roles span hypothesis generation, causal inference, temporal trend analysis, and cross-domain policy synthesis. Each agent is assigned a specialized profile and may be powered by a domain-adapted LLM instance. The entire workflow is automated, providing a robust and extensible framework for conducting social-climate analysis with minimal human intervention. This architecture supports scalable, interpretable insights into the societal dimensions of climate change, facilitating interdisciplinary engagement with data from institutions such as the United Nations, IPCC, and World Bank.

\textbf{Planner}. The planning phase begins after the task is received from the \textit{User} agent (Agent 1), who defines the problem scope-for instance, setting climate mitigation objectives such as emission reduction targets. The task is then processed by the \textit{Climate Strategist} (Agent 2), who formulates high-level actions and connects to external tools such as scenario planners or modeling platforms. Next, the \textit{Policy Planner} (Agent 5) generates a detailed, multi-step execution path tailored to the problem context, such as simulating policy pathways or evaluating socioeconomic interventions. During this stage, the \textit{Dialogue Manager} (Agent 4) coordinates agent communication, determining which agent should contribute next, thereby ensuring coherent message flow across the multi-agent pipeline. This planning workflow draws from expert reasoning encoded in agent profiles and is supported by general-purpose and domain-specific capabilities of the underlying AI agents. The configuration enables structured hypothesis formulation and plan decomposition aligned with climate objectives and constraints (see Table~\ref{tab:climate_agents}).

\textbf{Assistant}.Following the plan initialization, a collaborative execution sequence is launched by core agents. The \textit{Climate Scientist} (Agent 3) contributes domain-specific hypotheses, such as identifying critical variables (e.g., carbon sinks, feedback loops) relevant to the proposed scenario. For computational tasks, the \textit{Data Modeler} (Agent 7) applies analytical tools to extract key insights from environmental datasets-such as emissions trends or regional temperature anomalies-while the \textit{Plot Interpreter} (Agent 9) analyzes and explains generated visualizations. Simultaneously, the \textit{Knowledge Retriever} (Agent 10) searches for supporting evidence in relevant climate databases and official reports, including sources from the United Nations and World Bank. These documents are evaluated by the \textit{Fact Checker} (Agent 11), who verifies retrieved content for consistency, correctness, and citation relevance. To finalize the process, the \textit{Code Developer} (Agent 8) converts the structured output into usable formats, such as CSV files or formatted reports, for downstream use or archiving. This structured collaboration ensures that each agent’s capabilities are leveraged for interpretability, traceability, and task modularity, in alignment with the framework's objectives of supporting climate-informed decision making (Table~\ref{tab:climate_agents}).

\begin{table}[H]
	\caption{Multi-agent framework for climate change analysis. Each agent contributes a specialized function in the reasoning pipeline.}
	\label{tab:climate_agents}
	\centering
	\small
	\begin{tabular}{@{}clp{5.5cm}@{}}
		\toprule
		\textbf{Agent \#} & \textbf{Agent Name} & \textbf{Role} \\
		\midrule
		1 & User & Initiates the task, sets objectives, and provides feedback. \\
		2 & Climate Strategist (LLM Agent) & Plans inputs and manages global strategy using external tools. \\
		3 & Climate Scientist (Domain Reasoning Agent) & Proposes hypotheses from climate science domains. \\
		4 & Dialogue Manager (Coordinator Agent) & Coordinates message flow and agent turn-taking. \\
		5 & Policy Planner (Planning Agent) & Designs simulation-based policy plans. \\
		6 & Critic (Evaluation Agent) & Reviews plans for feasibility and climate alignment. \\
		7 & Data Modeler (Statistical Analysis Agent) & Extracts insights from climate and environmental data. \\
		8 & Code Developer (Code Generation Agent) & Implements data processing and visualization scripts. \\
		9 & Plot Interpreter (Visualization Analysis Agent) & Interprets graphical outputs and projections. \\
		10 & Knowledge Retriever (Retrieval Agent) & Gathers information from reports and databases. \\
		11 & Fact Checker (Verification Agent) & Validates and verifies cited content. \\
		\bottomrule
	\end{tabular}
\end{table}

\section {Results and discussion}

\subsubsection{Perceptions layer}

Agent planner:  Utilizing a text classification approach, the LLM-generated summaries collectively underscore the multifaceted nature of climate change prediction and the interdisciplinary efforts aimed at improving the accuracy, relevance, and interpretability of forecasting models. This method addresses both technical limitations and broader socio-political implications. Key thematic categories identified include variation in model frameworks, regional and global climate focus, equity considerations in climate vulnerability and adaptation, the granularity and scope of data across temporal scales, factors driving climate variability, performance metrics of forecasting techniques, and the strategic purposes of prediction-ranging from early-warning systems to long-term policy planning (Figure~\ref{fig:classification}, Appendix Figure~\ref{fig:appendix_classification}).

\begin{figure}[H]
	\centering
	\includegraphics[width=0.7\linewidth]{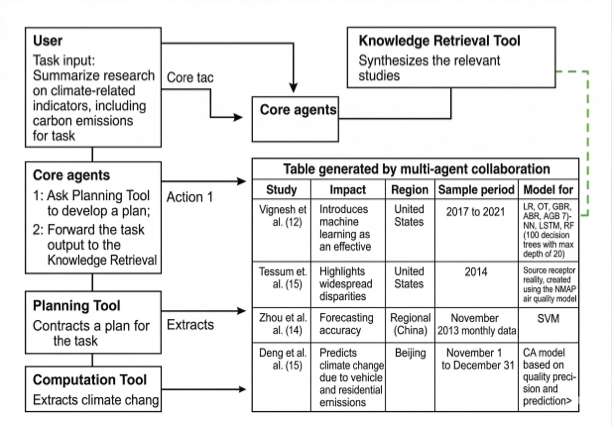}
	\caption{Multi-agent collaboration for synthesizing climate-related research. The user prompts the system to summarize studies on climate indicators such as carbon emissions. Core agents coordinate task planning, knowledge retrieval, and data synthesis. The planning tool generates a stepwise plan, while computation and retrieval agents extract relevant studies. The resulting table summarizes key features of the retrieved literature, including study impact, sample period, region, and modeling techniques.}
	\label{fig:screenshot004}
\end{figure}

\subsubsection{Reasoning layer}

\begin{figure}[H]
	\centering
	\includegraphics[width=0.7\linewidth]{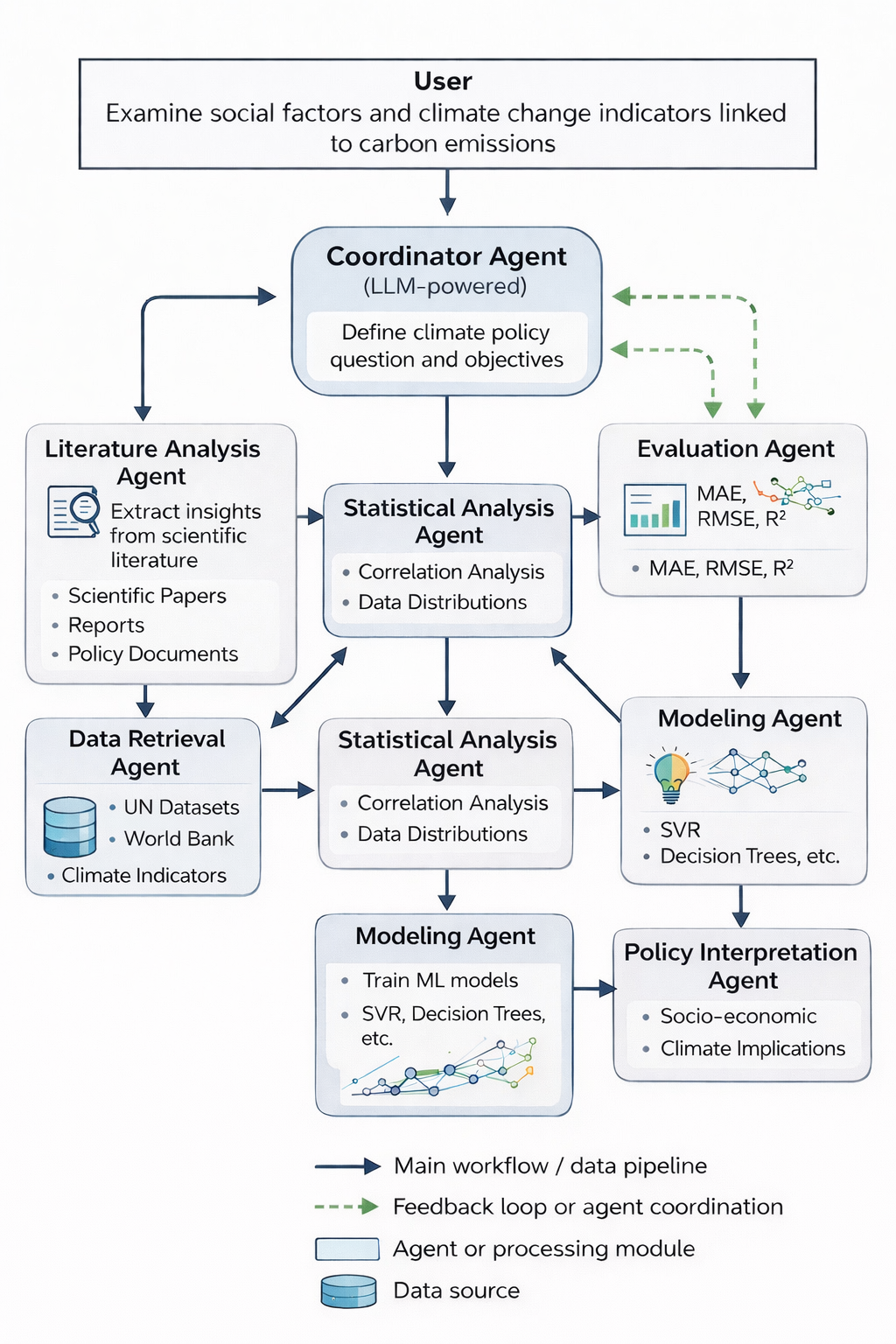}
	\caption{Reasoning framework for examining social and climate indicators linked to carbon emissions. The user initiates the process, and the core LLM agent coordinates classification, feature correlation, model building, pattern analysis, and evaluation. Outputs are synthesized to support interpretation of policy-relevant insights.}
	\label{fig:screenshot005}
\end{figure}

\textbf{Correlation}. Figure~\ref{fig:screenshot005} illustrates the reasoning pipeline conducted by a large language model (LLM) core agent in response to a user query concerning the social and environmental drivers of carbon emissions. The user prompt initiates an LLM-driven process that synthesizes information from scientific literature using a text classification module. This process identifies key studies, including their impact, modeling techniques, and associated socioeconomic implications. Once relevant data are extracted, the system performs feature correlation analysis to explore relationships between variables. A correlation matrix is generated to reveal statistically significant associations among social and climate-related indicators. The core agent then fits multiple predictive models, such as support vector regression (SVR) and decision trees, to capture underlying trends and interactions in the data. Pattern recognition modules are employed to analyze data distribution and detect nonlinear structures or clusters within the dataset. These distributions help to refine model fitting and reveal structural properties of the social-climatic variables under study. Model outputs are evaluated using standard metrics such as Mean Absolute Error (MAE), Root Mean Square Error (RMSE), and the coefficient of determination ($R^2$). The results are passed to a policy interpreter module that contextualizes findings by linking them to broader socioeconomic implications-for example, identifying how social inequality or urbanization patterns contribute to carbon emission trends.

\textbf{Causation}. After constructing a fully connected causal graph (See apendices 3), CAM pruning is applied to remove spurious edges and refine the structure. Variable selection retains important confounders, and correlation analysis eliminates irrelevant variables. This ensures that pruning does not introduce residual confounding. The final causal graph is validated using domain expertise. CAM pruning complements, but does not replace, confounding control methods, and may be used alongside expert-informed techniques such as LLMs.

This study adapts causal modeling techniques from Rolland et al. \cite{rolland2022score} where each variable is modeled as a function of its direct causal parents plus additive noise. The causal structure is inferred by identifying leaf nodes based on the variance of partial derivatives of the score function. Specifically, a node is a leaf if the variance of its self-derivative is zero, and it has a parent if cross-derivative variance is non-zero. Nodes are removed iteratively to determine a topological order. The score function is estimated using the Stein gradient estimator with ridge RBF kernel regression.

After constructing a fully connected causal graph, CAM pruning is applied to remove spurious edges and refine the structure. Variable selection retains important confounders, and correlation analysis eliminates irrelevant variables. This ensures that pruning does not introduce residual confounding. The final causal graph is validated using domain expertise. CAM pruning complements, but does not replace, confounding control methods, and may be used alongside expert-informed techniques such as LLMs .

This reasoning chain highlights the system’s ability to move from unstructured textual data to structured inference, combining statistical modeling with policy-aware interpretation. The integrated pipeline supports interdisciplinary insight into climate change by connecting data-driven analysis with human-centric questions about justice, equity, and policy outcomes.

\subsection{Operation layer}

Using the multi-agent framework, a comprehensive analysis was conducted on clean fuel access and urbanization indicators based on a diverse set of AI agents-generated questions. The agents collaboratively contributed domain-specific insights across understanding variables, analyzing historical data, modeling future scenarios, and evaluating policy impacts. A representative example of the system’s reasoning capabilities in response to a user query is illustrated in Figure~\ref{fig:llm_policy_response}.

As shown in Supplementary Figure~\ref{fig:screenshot007}, agent behaviors differ based on prompt strategies and align with varying levels of reflexive reasoning. In examining variable significance, the system identified key indicators as essential measures for assessing disparities in access to clean energy and tracking patterns of urban development. These variables provide critical context for understanding how demographic and infrastructural shifts influence environmental outcomes. The historical data analysis revealed consistent trends in fuel access and urbanization over time. Patterns observed across regions were shaped by major policy initiatives, technological advancements, and broader socioeconomic transitions. These insights underscore the importance of contextualizing data within historical and geopolitical developments. In the domain of predictive modeling, the system employed forecasting techniques to explore potential future trajectories of these variables. Scenario-based models highlighted plausible developments under varying conditions, showing strong interdependence between clean fuel accessibility, urban growth, and policy choices. Finally, the policy impact evaluation demonstrated how specific interventions have influenced observed trends. The analysis emphasized the effectiveness of integrated strategies that align clean energy initiatives with urban planning objectives. These approaches were shown to be instrumental in driving equitable environmental progress and in shaping sustainable futures.

Overall, the multi-agent collaboration enabled a layered, data-informed perspective that linked scientific understanding, predictive modeling, and policy design into a unified analysis framework. The results offer actionable insights that support evidence-based decision-making in climate and energy planning.

\begin{figure}[ht]
	\centering
	\fbox{
		\begin{minipage}{0.9\textwidth}
			\textbf{User:} \\
			Rural areas (\texttt{EG.CFT.ACCS.RU.ZS}) have lower access to clean fuels than urban ones (\texttt{EG.CFT.ACCS.UR.ZS}), and urbanization (\texttt{SP.URB.TOTL.IN.ZS}) is increasing.\\
			Could variability in clean fuel access affect the effectiveness of environmental policies at reducing emissions? \\[1em]
			
			\textbf{AI Social-Scientist :}
			\begin{itemize}
				\item \textbf{Policy Effectiveness:} Greater variability (standard deviation) in access weakens uniform policy outcomes by reflecting infrastructure inequity.
				\item \textbf{Emission Impact:} Urban-focused policies may yield quick results, but lasting success requires rural inclusion.
				\item \textbf{Urbanization Factor:} Urbanization shifts where and how policies work; less urbanized regions need tailored approaches.
				\item \textbf{Policy Design:} Reducing variability in access could lead to more equitable and impactful environmental strategies.
			\end{itemize}
			\vspace{0.5em}
			\textbf{Implications:}
			\begin{itemize}
				\item Targeted investments in rural infrastructure could improve long-term emission reductions.
				\item Adaptive policy frameworks should consider regional disparities in energy access.
				\item Integration of socio-environmental data with LLM reasoning may improve decision-making in complex systems.
			\end{itemize}
		\end{minipage}
	}
	\caption{AI Agents-generated response to a user question about variability in clean fuel access and its impact on environmental policy effectiveness. The dialogue illustrates how AI Agents can support reasoning over socio-environmental systems through hypothesis generation and contextual insight.}
	\label{fig:llm_policy_response}
\end{figure}

\subsection{Agent-based  evaluation}

The evaluation was conducted by AI agents following the published guidelines for the Stanford Agentic Reviewer.  The official scoring rubric assesses papers on seven dimensions—originality, importance of the research question, support of claims, soundness of experiments, clarity of writing, value to the research community, and contextualization relative to prior work—and then combines these scores to derive an overall rating.

\begin{figure}[ht]
	\centering
	\includegraphics[width=0.7\linewidth]{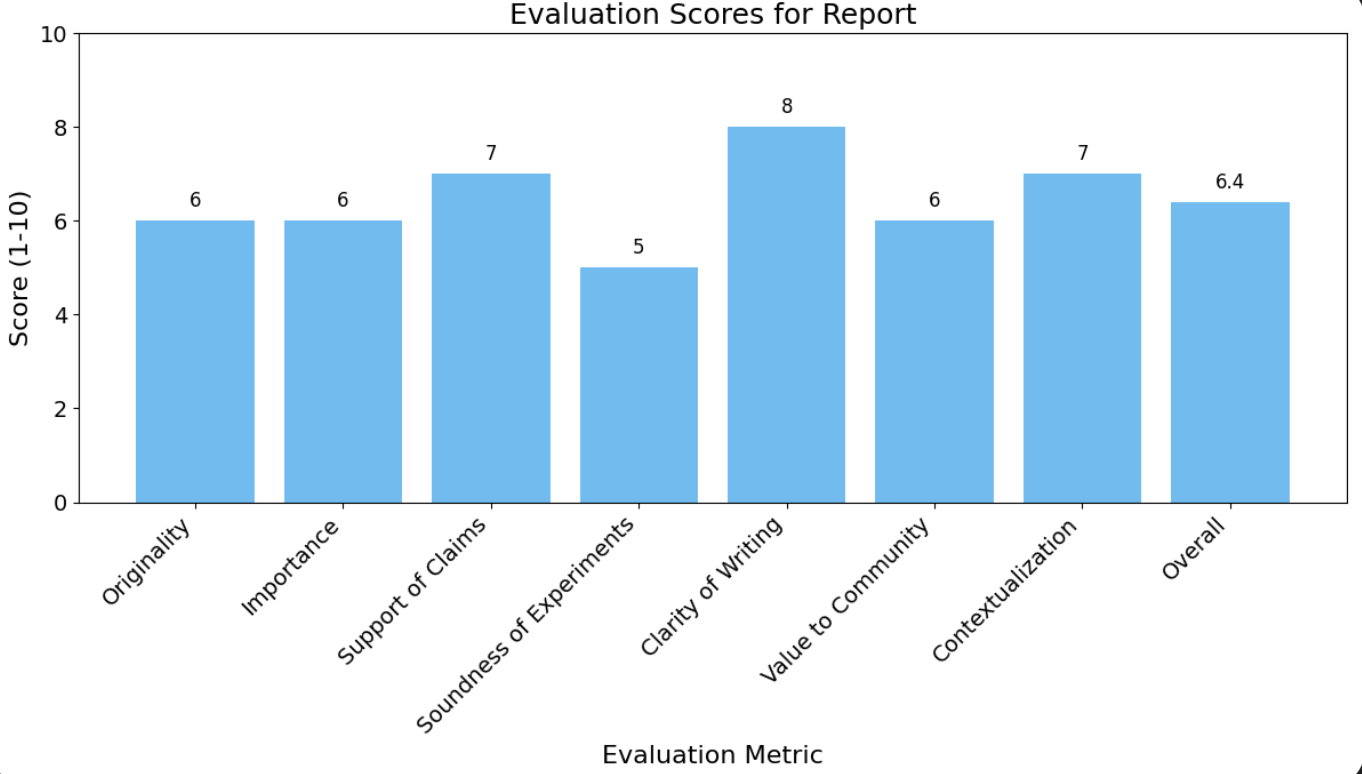}  
	\caption{Agent-based evaluation scores across the seven Agentic Reviewer criteria and the resulting overall score.  Higher scores indicate stronger performance in that dimension.}
	\label{fig:evaluation_scores}
\end{figure}

Figure~\ref{fig:evaluation_scores} shows a bar chart illustrating the scores across each criterion defined by the Stanford Agentic Reviewer.  These scores reflect the assessment of the report produced by \textit{ClimateAgent}, summarizing its originality, importance of the research question, support of claims, soundness of experiments, clarity of writing, value to the research community, contextualization relative to prior work, and the resulting overall score.

\begin{figure}[ht]
	\centering
	\includegraphics[width=0.7\linewidth]{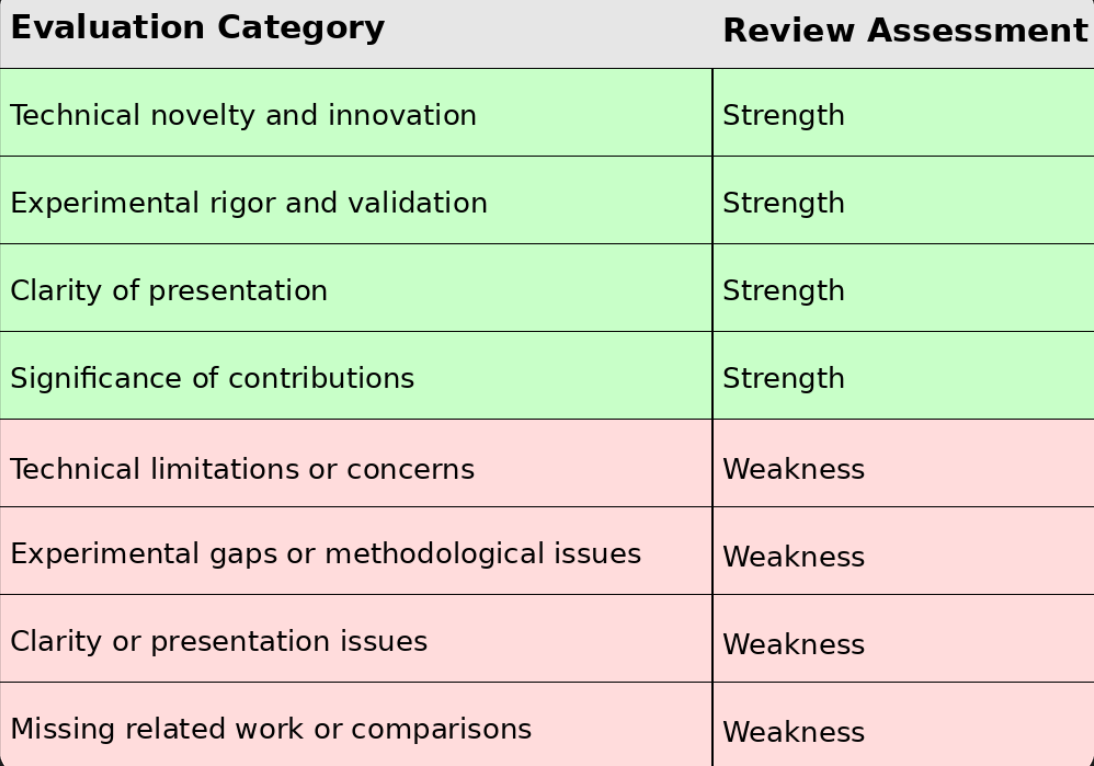}  
	\caption{Matrix summarising the Agentic Reviewer feedback.}
	\label{fig:evaluation_matrix}
\end{figure}

Figure~\ref{fig:evaluation_matrix} presents a matrix summarizing the feedback from the Stanford Agentic Reviewer.  This matrix summarizes the agent-based evaluation, where the Stanford Agentic Reviewer assesses the generated work across key research criteria, identifying strengths in novelty, experimental rigor, clarity, and contribution significance, while noting weaknesses related to technical limitations, experimental gaps, presentation details, and missing comparisons with prior work. The agent-based evaluation scores 6 for originality and importance, 7 for support of claims and contextualization, and  8 for clarity of writing, indicating clear presentation and reasonably supported arguments. However, soundness of experiments receives the lowest score of 5, suggesting the need for stronger experimental validation. The overall score of 6.4 reflects a solid but improvable contribution.

\section{Limitations and Future Perspectives}

While this study demonstrates the potential of multi-agents  systems in reasoning about socio-climatic issues, several limitations remain. Drawing from Minsky’s proposition that intelligent agents can emerge through the coordination of simpler components, it becomes clear that even advanced multimodal systems do not yet achieve true awareness. Although such systems enhance observable reasoning capabilities, awareness-particularly in tasks involving causal inference and moral understanding-remains fundamentally out of reach for current models.

The evaluation presented in this study touches upon the boundaries of causal reasoning and inference within language models. Yet, posing meaningful causal questions to agents requires more than technical design; it also involves acknowledging the social norms and moral underpinnings embedded in human-machine communication. These dimensions-social, cultural, and ethical-are far more complex than the statistical language patterns on which these models are based \cite{van2015detection,wang-etal-2018-modeling,forbes-etal-2020-social,cui2024odysseycommonsensecausalityfoundational}.

One significant limitation lies in the framework’s dependency on input data quality. The accuracy and reliability of the causal inferences generated are directly tied to the completeness and representativeness of the underlying datasets. Inadequate or biased data can lead to misleading conclusions and may adversely affect policy relevance. In addition, the reasoning structure adopted in this framework relies on assumptions derived from global carbon emissions and existing climate models. These assumptions may not hold across different regions or socio-economic contexts, potentially restricting the framework’s flexibility when applied to diverse scenarios.

Another limitation concerns the generalizability of the results. The framework is optimized for specific climate-related tasks and may not transfer seamlessly to unrelated domains without significant adaptation. Moreover, the interpretability of outputs may vary depending on task complexity and domain-specific knowledge embedded in the agents' profiles.

Looking forward, future research should aim to incorporate broader and more heterogeneous data sources, especially those rooted in localized social realities. Strengthening the integration between technical reasoning and real-world social dynamics could significantly enhance the system’s policy relevance. Expanding the framework into adjacent verticals-such as education, public health, and urban resilience-would also increase its practical scope and impact. Furthermore, formalizing causal understanding within AI agents-based systems remains a critical research frontier. Approaches that blend symbolic reasoning, simulation, and ethical modeling may help overcome current limitations. As AI agents evolve toward more awareness-like behavior, developing robust interpretability and accountability frameworks will be essential to ensure that their insights are not only useful but also socially and ethically grounded.

\section*{References}

\bibliographystyle{plainnat}
\bibliography{sample}

\appendix

\section*{Supplementary Materials}

\begin{figure}[H]
	\centering
	\includegraphics[width=0.5\linewidth]{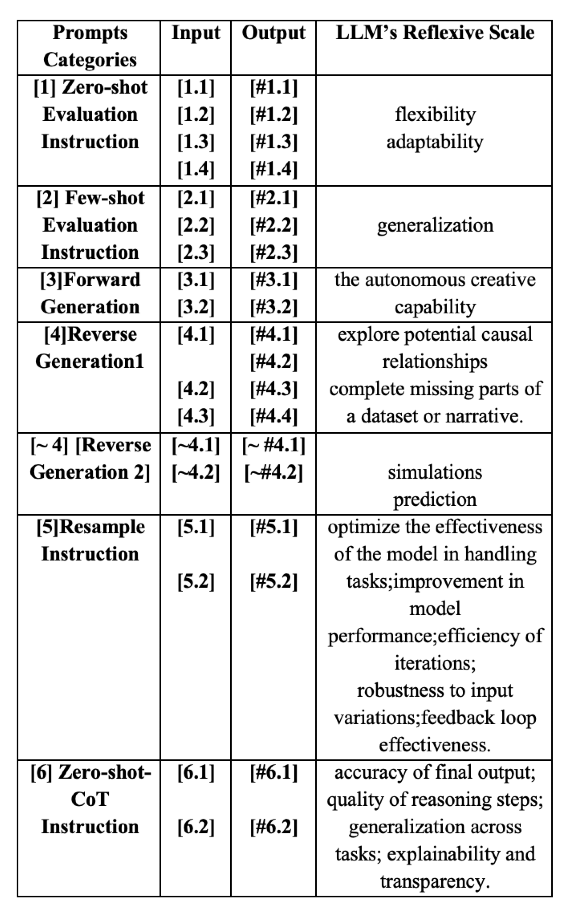}
	\caption{Agents’ operation modes across prompt categories and their alignment with LLM’s reflexive reasoning scale. This matrix illustrates how different prompt strategies (e.g., zero-shot, few-shot, generative) correspond to specific input-output patterns and reflect varying cognitive capacities such as adaptability, generalization, and transparency.}
	\label{fig:screenshot007}
\end{figure}

\subsection*{Image Generation Method}

All illustrative figures depicting system architecture and reasoning flows were generated using LLM-guided image generation via the \texttt{image\_gen.text2im} command. This method utilizes natural language prompts to synthesize clean, vector-style infographics. The generation process ensures clarity, role labeling, and directional flow for visualizing complex multi-agent interactions.

\begin{verbatim}
	DALL·E-style prompt (LLM command)
	
	image_gen.text2im({
		"prompt": "A flowchart diagram titled 'ClimateAgents' showing a multi-agent system involving GPT-4o and GPT-4-turbo agents. At the center, there are three vertically aligned hexagons labeled: 'Core memory', 'Social scientist (GPT-4o)', and 'Climate scientist (GPT-4o)'. To the left of each are rectangular tool boxes: 'Planning tool' with the description 'Defines a set of steps to achieve the desired target', and 'Literature review tool' with 'Synthesizes relevant findings from research'. To the right are three other tool boxes labeled 'Policy planner (GPT-4-turbo)' with 'Creates detailed policy plans', 'Data modeler (GPT-4-turbo)' with 'Builds predictions', and 'Plot interpreter (GPT-4o)' with 'Analyzes generated visual output'. Red arrows labeled 'query' go from the central hexagons to the tools, and green arrows labeled 'response' go from tools back to the central roles. Use a clean, modern vector infographic style with consistent blue tones for roles, and red/green arrows for flows.",
		"size": "1024x1024"
	})
	
	
	DALL·E-style prompt (LLM command)
	
	image_gen.text2im({
		"prompt": "A diagram titled 'Layered architecture for multimodal-agent reasoning systems'. It features three distinct layers: (1) A yellow 'Perception Layer' box on the left, labeled with text describing how multimodal inputs like text, tables, and images are structured and interpreted. (2) A blue 'Reasoning Layer' box in the center-right, powered by a large circle labeled 'LLM', describing tasks like planning, inference, and contextual reasoning. (3) A green 'Operation Layer' box on the bottom-left, connected with arrows, explaining actions such as literature retrieval, data processing, and visual analysis. Each layer includes small icons representing the tasks. The diagram has arrows showing data flow between layers, using a clear and professional infographic style.",
		"referenced_image_ids": ["2a605bdc-f679-48bf-bd68-e95d3e571e38"]
	})
\end{verbatim}

*These prompts were executed via OpenAI's multimodal command interface to generate consistent, modern infographics for illustrating model architecture and collaborative agent pipelines.

\subsection{Method Summary}

This study adapts causal modeling techniques from Rolland et al.~(2022), where each variable is modeled as a function of its direct causal parents plus additive noise. The causal structure is inferred by identifying leaf nodes based on the variance of partial derivatives of the score function. Specifically, a node is a leaf if the variance of its self-derivative is zero, and it has a parent if cross-derivative variance is non-zero. Nodes are removed iteratively to determine a topological order. The score function is estimated using the Stein gradient estimator with ridge RBF kernel regression.

\subsection{Causal Graph Construction}

Given a dataset with 16 variables, \(X_1, \ldots, X_{16}\), the data-generating model assumes:
\[
X_i = f_i(\text{pa}_i(X)) + \epsilon_i, \quad i = 1, \ldots, 16,
\]
with additive noise \(\epsilon_i\), and joint distribution:
\[
p(x) = \prod_{i=1}^{16} p(x_i \mid \text{pa}_i(x)), \quad \log p(x) = \sum_{i=1}^{16} \log p(x_i \mid \text{pa}_i(x)).
\]
The score function is defined as \(s(x) = \nabla \log p(x)\). A node \(j\) is a leaf if:
\[
\text{Var}_X\left(\frac{\partial s_j(X)}{\partial x_j}\right) = 0,
\]
and it has a parent \(i\) if:
\[
\text{Var}_X\left(\frac{\partial s_j(X)}{\partial x_i}\right) \neq 0.
\]
Using these conditions, nodes are sequentially removed to establish a topological order. The score Jacobian is estimated via Stein gradient with ridge RBF kernels.

\newpage

\end{document}